\documentclass[twocolumn,prl] {revtex4}

\ifx\pdfoutput\undefined
  \usepackage[dvips]{graphicx}
\else
  \usepackage[pdftex]{graphicx}
\fi

\usepackage{dcolumn}
\usepackage{amsmath}
\usepackage{latexsym}

\begin{document}
\title{Metallic single-electron transistor without traditional
tunnel barriers}
\author{V.A. Krupenin$^{\text{1}}$, A.B. Zorin$^{\text{2,3}}$, D.E. Presnov$^{\text{2}}$,
M.N. Savvateev$^{\text{1}}$, and J. Niemeyer$^{\text{3}}$}
\affiliation{$^{\text{1}}$Laboratory of Cryoelectronics, Moscow
State University,
119899 Moscow, Russia\\
$^{\text{2}}$Nuclear Physics Institute, Moscow State University,
119899 Moscow,Russia\\
$^{\text{3}}$Physikalisch-Technische Bundesanstalt, 38116
Braunschweig, Germany}

\begin{abstract}
We report on a new type of single-electron transistor (SET)
comprising two highly resistive Cr thin-film strips
($\sim$1~$\mu$m long) connecting a 1~$\mu$m-long Al island to two
Al outer electrodes. These resistors replace small-area oxide
tunnel junctions of traditional SETs. Our transistor with a total
asymptotic resistance of 110~k$\Omega$ showed a very sharp Coulomb
blockade and reproducible, deep and strictly e-periodic gate
modulation in wide ranges of bias currents $I$ and gate voltages
$V_g$. In the Coulomb blockade region ($|V| \leq$ about 0.5~mV),
we observed a strong suppression of the cotunneling current
allowing appreciable modulation curves $V(V_{g})$ to be measured
at currents $I$ as low as 100~fA. The noise figure of our SET was
found to be similar to that of typical Al/AlO$_{x}$/Al
single-electron transistors, viz. $\delta Q \approx 5\times
10^{-4}e/\sqrt{\rm{Hz}}$ at 10~Hz.
\end{abstract}
\maketitle

\section{Introduction}

The single-electron  transistor (SET) is a system of two
ultra-small metal-insulator-metal tunnel junctions attached to a
small island which is capacitively coupled to a gate electrode.
Due to their considerable resistance, $R \gg R_Q \equiv h/4e^2
\cong 6.5$~k$\Omega$, the tunnel junctions ensure charge
quantization on the island. On the other hand, the junctions still
enable the (correlated) charging and discharging of the island by
individual electrons when the temperature is sufficiently low,
$k_BT \ll E_c$. Here $E_c = e^2/2C_{\Sigma}$ is the charging
energy and $C_{\Sigma}=C_1+C_2+C_0+C_g$ the total capacitance of
the island which includes the capacitances of the junctions
$C_{1,2}$, the self-capacitance of the island $C_0$ and the
capacitance between the island and gate electrode $C_g$.
Transport of electrons is controlled by the transistor gate
polarizing the island and therefore changing the Coulomb blockade
threshold. Increase in the gate voltage $V_g$ causes a stepping
increment of the number of electrons on the island and this leads
to e-periodic dependence of the $I-V$ characteristic on $V_g$. Due
to this effect, the transistor provides a means for measuring the
polarization charge on its island with sub-electron accuracy.
This property of SET was successfully exploited in many
experiments to measure and monitor sub-electron quantities of
charge in mesoscopic systems (see some examples in Refs.
\cite{box,pump,trap1,trap2,QHall}). Different materials and
methods have been used for the fabrication of SETs. Although
substantial progress has been achieved in the fabrication
techniques, there is still a demand for devices which have good
electric parameters, are easy to fabricate and less subject to
electrical shock destruction. In this paper we present a new type
of SET electrometer having a potential to meet these
requirements. The idea of our work was to fabricate and
characterize a metallic transistor with high-ohmic ($R \gg R_Q$)
Cr-film microstrips replacing the traditional (oxide) tunnel
barriers between island and outer electrodes. The junctions of
this type should, in our opinion, also ensure sufficient
isolation of the island, leading to correlated electron motion
across the device \cite{Naz}. Below we report on the single
electron characteristics of our device.

\section{Sample Fabrication}

The structures were fabricated by shadow evaporation through a
suspended mask \cite{Fult} on Si substrate buffered by an
AlO$_{x}$ layer (about 200~nm thick). Since we aimed to obtain
reliable metallic contacts between Cr and Al films, the
fabrication process was carried out $in$ $situ$ in three steps.
First, the Al film 10~nm thick was deposited by e-gun at the
first angle. It formed the bottom layers of the island and outer
electrodes. Secondly, the Cr film $6\div8$~nm in thickness  was
evaporated at a small residual oxygen pressure
($\approx10^{-5}$~mbar). This evaporation was performed at another
angle to overlap the bottom Al layer and give rise to the
electrical connection between the island and outer electrodes. The
nominal contact area between the Al and Cr layers was 100~nm by
120~nm. Finally, the second Al layer (about 30~nm thick) was
evaporated at the first angle so that the ends of the Cr strips
were clasped by Al electrodes from the bottom and from the top.
The gate electrode, completing the transistor structure, obtained
by evaporation of both Al and Cr films, and it was positioned near
the island. The top view of the resulting structure is presented
as a sketch in Fig.~1 where the stray metal shadows are not shown
for the sake of clarity.

\begin{figure}
\begin{center}
\leavevmode
\includegraphics[width=3in]{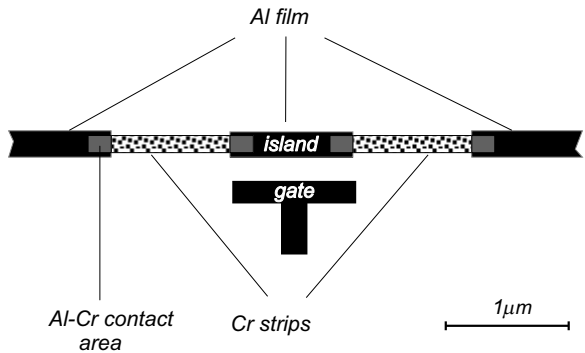}
\caption {The geometry (top view) of the Cr-film transistor
structure. For clarity, the width of the strips (about 100~nm) is
shown somewhat larger.}
\label{fig:fig1}
\end{center}
\end{figure}

\section{Characterization of chromium strips}

Before studying the transistor structures we had characterized
the detached Cr strips (1.0$\div$1.4~$\mu$m long, 100~nm wide and
$6\div8$~nm thick) identical to those used in the transistor
samples. In order to check a possible influence of the electric
field on the electron transport in Cr strips, each strip was
supplied by an individual gate electrode placed nearby. The Cr
strips had different resistances per square $R_{\Box}$ (from
$2.5~$k$\Omega$ to $ 14~$k$\Omega$) depending on their thickness.
The $I-V$ curves of these strips measured at $T=30$~mK are
presented in Fig. 2.

\begin{figure}[h]
\begin{center}
\leavevmode
\includegraphics[width=3in]{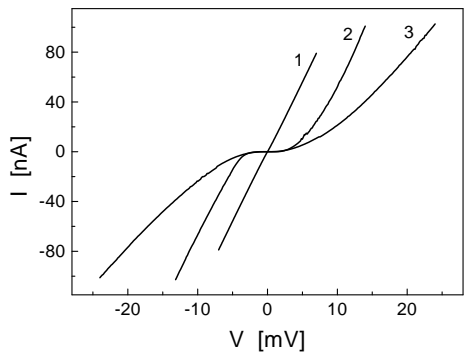}
\caption {$I-V$ curves of $6\div8$~nm thick Cr strips having
different resistances per square $R_{\Box}$. Curve 1:
1~$\mu$m-long low resistance strip,
$R=25$~k$\Omega~(R_{\Box}\approx2.5$~k$\Omega),~V_{off}\approx70~\mu$V;
Curve 2: 1.4~$\mu$m-long medium resistance strip,
$R=80$~k$\Omega~(R_{\Box}\approx
5.5$~k$\Omega),~V_{off}\approx5$~m$V$; Curve 3: 1~$\mu$m-long
high resistance strip, $R=140$~k$\Omega~ (R_{\Box}\approx
14$~k$\Omega),~V_{off}\approx8$~mV.}
\label{fig:fig2}\end{center}\end{figure}

The low-resistive sample 1 showed insignificant nonlinearity of
its $I-V$ curve, while the samples 2 and 3 (with $R_{\Box}\approx
R_Q$ or higher) exhibited a clear blockade region of current in
their $I-V$ curves. The offset voltage $V_{off} \approx 5-8$~mV
was approached at $V\approx10-15$~mV. We associate such
characteristics of the samples with intrinsic Coulomb blockade
effects (partially depressed by quantum fluctuations) in small Cr
granules forming the strips. When a tunnel resistance of thin
barriers separating the metallic granules increases up to the
threshold value about $R_Q$, the array of these granules
gradually turns to the blockade state. At lower resistances of the
barriers, the quantum fluctuations completely wash out the
charging effect. Assuming a uniform two-dimensional arrangement
of the granules we can directly relate the sheet resistance
$R_{\Box}$ to the intergranular barrier resistance.

We made an attempt to use the gates to control the transport
current through the strips, but the $I-V$ curves did not show
noticeable dependence on the gate voltage in the significant
range $|V_g| \leq 1$~V. This property can also be understood on
the basis of the two-dimensional array model: since the background
charges $-e/2 \leq Q_{0i} \leq e/2$, $i=1,2,...,N$ are randomly
distributed over the large number $N$ of granular islands, their
additional polarization ($\propto V_g$) produced by the common
gate does not change the random character of their distribution.
Therefore the $I-V$ curve remained almost the same when value of
$V_g$ changed. (Note that for the same reason, the effect of $V_g$
on the $I-V$ curve of a one-dimensional array of 13 tunnel
junction \cite{Kuz}, \cite{Dels} was not periodic and not as
strong as in the case of transistor samples.)

\section{Characterization of the transistor}

Our transistor structure comprised two sections of Cr-film strips
with a sheet resistance $R_{\Box}\approx4$~k$\Omega$. The total
asymptotic resistance $R$ of the sample was about 110~k$\Omega$.
This value is in fair agreement with the resistance of two Cr
sections connected in series, $m R_{\Box} \approx104$~k$\Omega$,
where $m \approx 26$ is the total number of squares. This meant
that the resistances of possible tunnel junctions at the Al-Cr
interfaces were seemingly small ($< R_Q$). The resistance $R$ as
well as the shape of the transistor $I-V$ curves were practically
insensitive to the perpendicular magnetic field up to the strength
$B=1$~T. (For that reason, most of the measurements were performed
in a zero magnetic field.) At low temperature, $T \approx 30$~mK,
this value of $B$ ensured the normal state of Al and hence the
normal density of states in the electrodes. The fact that the
$I-V$ curves remained the same as those for superconducting Al
electrodes substantiates our conclusion regarding the negligible
role of the Al-Cr interface resistance.

\begin{figure}
\begin{center}
\leavevmode
\includegraphics[width=3in]{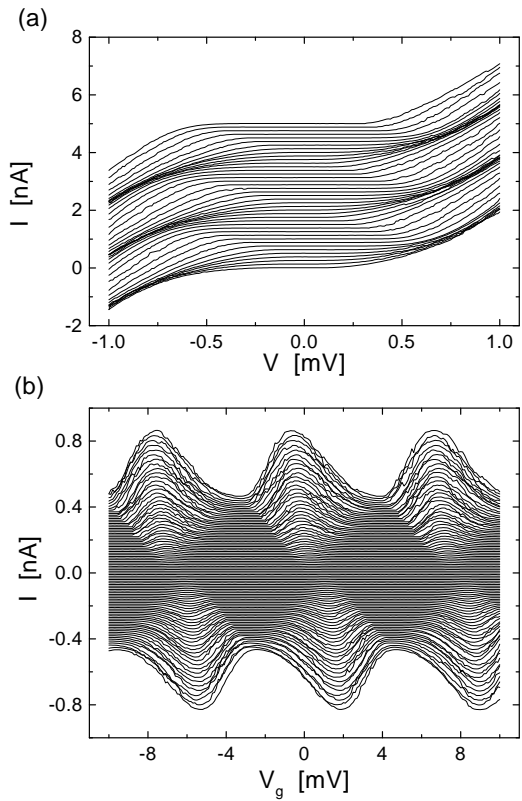}
\caption {Experimental $I(V,V_g)$ surfaces of the Cr-film
transistor at 30~mK: (a) $I-V$ curves measured at
$V_{g}=0\div20$~mV with a step of 0.5~mV; (b) $I-V_g$ curves
measured at $V=-0.45\div0.45$~mV with a step of 0.01~mV.}
\label{fig:fig3}\end{center}\end{figure}

The current-voltage and gate-modulation curves of our transistor
are presented in Fig.~3. These characteristics clearly
demonstrate the typical single-electron behavior of the device:
strong blockade effect (maximum size of blockade region $V_{max}
\approx 0.5$~mV) and large modulation of the blockade by the
gate. The zero bias resistance of the sample in the "open" state
is measurable, about 60~M$\Omega$, and drastically decreases with
the increase in $V$. The modulation curves are strictly periodic
with a period $\Delta V_g \approx 7$~mV. The constant amplitude
and stable period persist over the whole range of measurements,
viz. 40 periods.

From the characteristics obtained we conclude that the gate
polarizes the island via the capacitance $C_g = e/\Delta V_g
\approx 23$~aF. This value agrees with the sizes and mutual
arrangement of island and gate. As long as the gate had a minor
effect on the blockade in the detached strips, there is good
reason to believe that the transistor gate  does not change the
conductance and the offset owing to Cr lines. Thus, the device
operates in a similar way as the usual SET with tunnel barriers:
the transport is realized by means of single electrons which
arrive at (depart from) the island through highly resistive
strips. The offset voltage of the transistor
$V_{off}\approx1.1$~mV was evaluated from the asymptotic
characteristics of the $I-V$ curve taken at large $V>3$~mV.

Having evaluated voltage-to-charge ratio d$V/$d$Q$ from the
periodic dependence $V$ versus $V_g$, we measured noise spectra
at different working points (on the slope, in the maximum and
minimum) of the modulation curve at $I=5$~pA. The data (the
strongest noise on the slope) clearly pointed to the charge
nature of the dominant noise. The shape of the spectrum was close
to 1/$f$ and the charge noise level at 10~Hz was evaluated as
$5\times10^{-4}e/\sqrt{\rm{Hz}}$. The value obtained is also
typical of Al/AlO$_{x}$/Al transistors with the island size to be
about 1~$\mu$m. We suppose that the appreciable noise in our
structure is determined by the considerable contact area between
island and substrate, and the figure can be improved by
decreasing the sizes of the island \cite {JLowT}.

\section{Discussion of results}

We found several remarkable features of the transistor sample
investigated. In contrast to the property of traditional SETs, we
observed significant nonlinearity of the $I-V$ curves in the open
state, namely, $I \propto V^{\alpha}$, where $\alpha \approx 2.5$
(cf. the linear behavior of the tunnel barrier transistor). In
the Coulomb blockade state the $I-V$ curve exhibits the large
region with vanishingly small current (see, e.g., Fig.~3b). The
current decay in the region of the blockade corner is
substantially steeper ($\alpha>10$) than the dependence $I \propto
V^3$ which is typical of tunnel barrier transistors at low $T$
\cite{Geerligs} because of the cotunneling effect \cite{Aver}.
This dependence indicates that the cotunelling process, i.e.
arrival of an electron at the island and simultaneous escape of
another electron from the island, is strongly suppressed. Such a
suppression of cotunneling occurs in one-dimensional arrays of
tunnel junctions with a large number of junctions \cite{Geerligs}
or in SETs supplied with miniature on-chip resistors, the
so-called R-SET \cite{R-SET}, due to the effect of a dissipative
electromagnetic environment \cite{Odints} created by these
resistors. The latter system has the advantage to efficiently
control the transport current by gate. It is interesting that -
although our device is based on another junction type - its
behavior is similar to that of R-SET, i.e. it combines the
property of deep modulation of single electron current and
considerable suppression of the cotunneling current.

\begin{figure}
\begin{center}
\leavevmode
\includegraphics[width=3in]{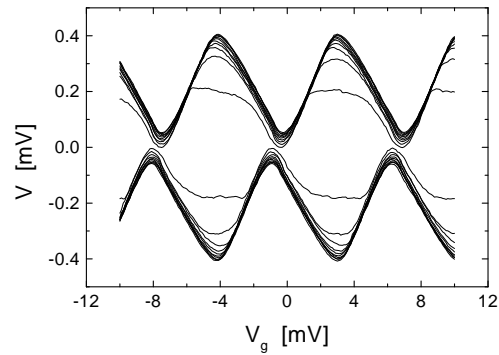}
\caption {$V-V_{g}$ modulation curves of Cr-film transistor
measured for bias currents $I=0.1,~0.2,~...1$~pA at $T$=30~mK.}
\label{fig:fig4}\end{center}\end{figure}

Owing to the significant suppression of cotunneling, we were able
to observe the $V-V_g$ modulation characteristics feeding a very
small current $I$ through the transistor (see Fig.~4). For
example, at $I=I_0=200$~fA, the maximum value of $V$ approaches
the value $\approx 0.3$~mV. The evaluation of the cotunneling
current in the symmetric tunnel-barrier SET device having similar
parameters ($R=110$~k$\Omega$ and $V_{off}\approx1.1$~mV) at the
same $V=0.3$~mV yields the value $I_{cot} \approx 15$~pA, i.e.
about two (!) orders of magnitude larger than $I_0$.

\section{Summary.}
We demonstrated a new type of single-electron transistor with
highly resistive Cr film strips instead of the traditional tunnel
junction. Although a complete theory of its operation has not yet
been elaborated, the device apparently behaves like a single
electron system. It shows remarkable characteristics which are
qualitatively similar to those of SETs with usual tunnel
junctions. It is one of the advantages of our SET device that it
is relatively easy to fabricate. The main difficulty consists in
evaporating Cr film with a well-defined sheet resistance. The
investigated device consists of elements rather large in size as
compared to traditional tunnel junction SET devices. Therefore,
there apparently is room for increasing the operating temperature
of the metallic transistors without tunnel barriers by decreasing
the sizes of strips and island.

Strong suppression of the cotunneling current, reasonable
resistance in the open state and its insensitivity to the magnetic
field are promising indications that this device can do the job
of a low-leakage switch in delicate experiments with mesoscopic
structures like single electron traps, pumps, etc. Another
potential application of this transistor could be in digital SET
circuits especially if the problem of the background charge is not
as severe as in traditional SET devices. A further study of the
background charge noise and drift in our devices is therefore to
the foreground.

Finally, the metallic Cr-film transistors have a better tolerance
as regards electric shock destruction because they have no tunnel
barriers which are usually very sensitive to voltage pulses when
tested.

\section{Acknowledgments}

We wish to thank M.Yu.~Kupriyanov for valuable discussions. This
work was supported by the German BMBF, the Russian Fund for Basic
Research, the Russian Program ''Physics of Solid State
Nanostructures'' and the Russian Ministry of Science.

\end{document}